\documentclass[11pt,twoside]{article}

\usepackage{asp2006}
\usepackage{epsf}
\usepackage{psfig}
\usepackage{lscape}

\begin{document}
\title{Exploring the Impact of Galaxy Interactions over Seven Billion Years with CAS}

\vspace{-0.3 cm}

\author{
Sarah Miller (UT Austin), S. Jogee (UT Austin), C. Conselice (Nottingham), K. Penner (UT Austin), 
E. Bell (MPIA), X. Zheng (PMO), C. Papovich (Arizona), R. Skelton (MPIA), R. Somerville (MPIA),
H. Rix (MPIA), F. Barazza (EPFL), M. Barden (Innsbruck), A. Borch (MPIA), S. Beckwith (JHU), 
J. Caldwell (UT McDonald), B. Haeussler (Nottingham), C. Heymans (UBC IAP), K. Jahnke (MPIA), D. McIntosh (UMass), 
K. Meisenheimer (MPIA), C. Peng (NRC HIA STScI), A. Robaina (MPIA), S. Sanchez (CAHA), 
L. Wisotzki (AIP), C. Wolf (Oxford)}

\begin{abstract}
We explore galaxy assembly over the last seven billion years by characterizing ``normal'' 
galaxies along the Hubble sequence, against strongly disturbed merging/interacting galaxies 
with the widely used CAS system of concentration (C), asymmetry (A), and `clumpiness' (S) 
parameters, as well as visual classification. We analyze Hubble Space Telescope (HST) ACS 
images of $\sim$4000 intermediate and high mass ($\,M/\,M_\odot>10^9$) galaxies from the 
GEMS survey, one of the largest HST surveys conducted to date in two filters. 
We explore the effectiveness of the CAS criteria [$A>S$ and $A>$~0.35] in separating normal 
and strongly disturbed galaxies at different redshifts, and quantify the recovery and 
contamination rate. We also compare the average star formation rate and the cosmic star 
formation rate density as a function of redshift between normal and interacting systems 
identified by CAS.
\end{abstract}

\vspace{-0.7 cm}
\section{Introduction}
Galaxy mergers and interactions are believed to have a profound impact on the structural 
evolution and star formation activity of galaxies. The recent advent of large space-based 
surveys with thousands of galaxies, such as GEMS, GOODS, COSMOS, and AEGIS, has led to the 
use of different methods for identifying interacting/merging galaxies, such as detailed 
visual classification, and quantitative codes like CAS (Conselice et al. 2000). However, 
detailed assessment of the results from different methods has been altogether lacking.

\vspace{-0.3 cm}
\section{CAS}
We quantify the concentration, asymmetry, and high-spatial frequency clumpiness of the 
sample galaxies using the CAS concentration ($C$) (Bershady et al. 2000), asymmetry ($A$) 
and clumpiness ($S$) indices (Conselice et al. 2000). The concentration index ($C$) is 
the ratio of the two radii containing $80\%$ and $20\%$ of the total flux, which is then 
normalized logarithmically. The asymmetry index ($A$) is found by rotating the original 
image of the galaxy by 180 degrees, and then subtracting this image from the original. 
The residual flux of this subtracted image is then normalized by the original galaxy's 
flux. The clumpiness index ($S$) is computed by reducing the original galaxy's effective 
resolution to create a new image that is smoothed so that the high-frequency structure 
has been washed out. The original image is then subtracted from the smoothed image to 
produce a residual map, which then contains only the high frequency part of the original 
galaxy's light distribution. The flux of this light is then summed and divided by the sum 
of the original galaxy's flux to obtain the $S$ index. The CAS criteria for identifying 
distorted/interacting systems ($\,A>S$ and $\,A>0.35$) are empirically derived from nearby 
galaxies (Conselice et al. 2003).

\begin{figure}[!ht]
\plotfiddle{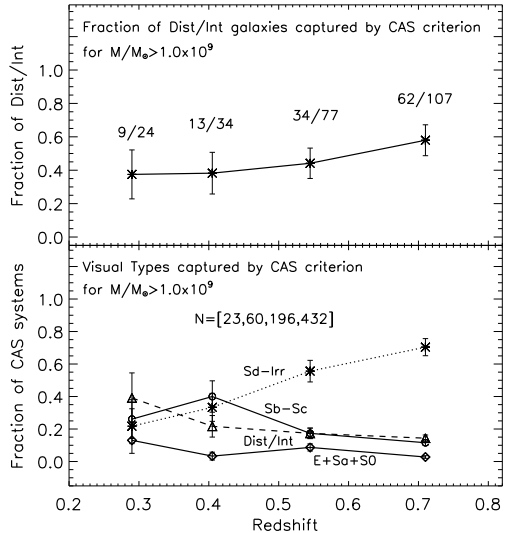}{190 pt}{0}{40}{40}{-110 pt}{-60 pt}
\caption{The top panel shows, by redshift, the fraction of visually-classified, strongly 
distorted/interacting systems that the CAS criteria also identify as distorted/interacting. 
The bottom panel shows which visually-classified systems CAS is identifying as distorted/
interacting. Notice the large fraction of Sd-Irr type galaxies as redshift increases, 
in agreement with Jogee et al. (2008).}
\end{figure}
\vspace{-0.3 cm}

\begin{figure}[!ht]
\plottwo{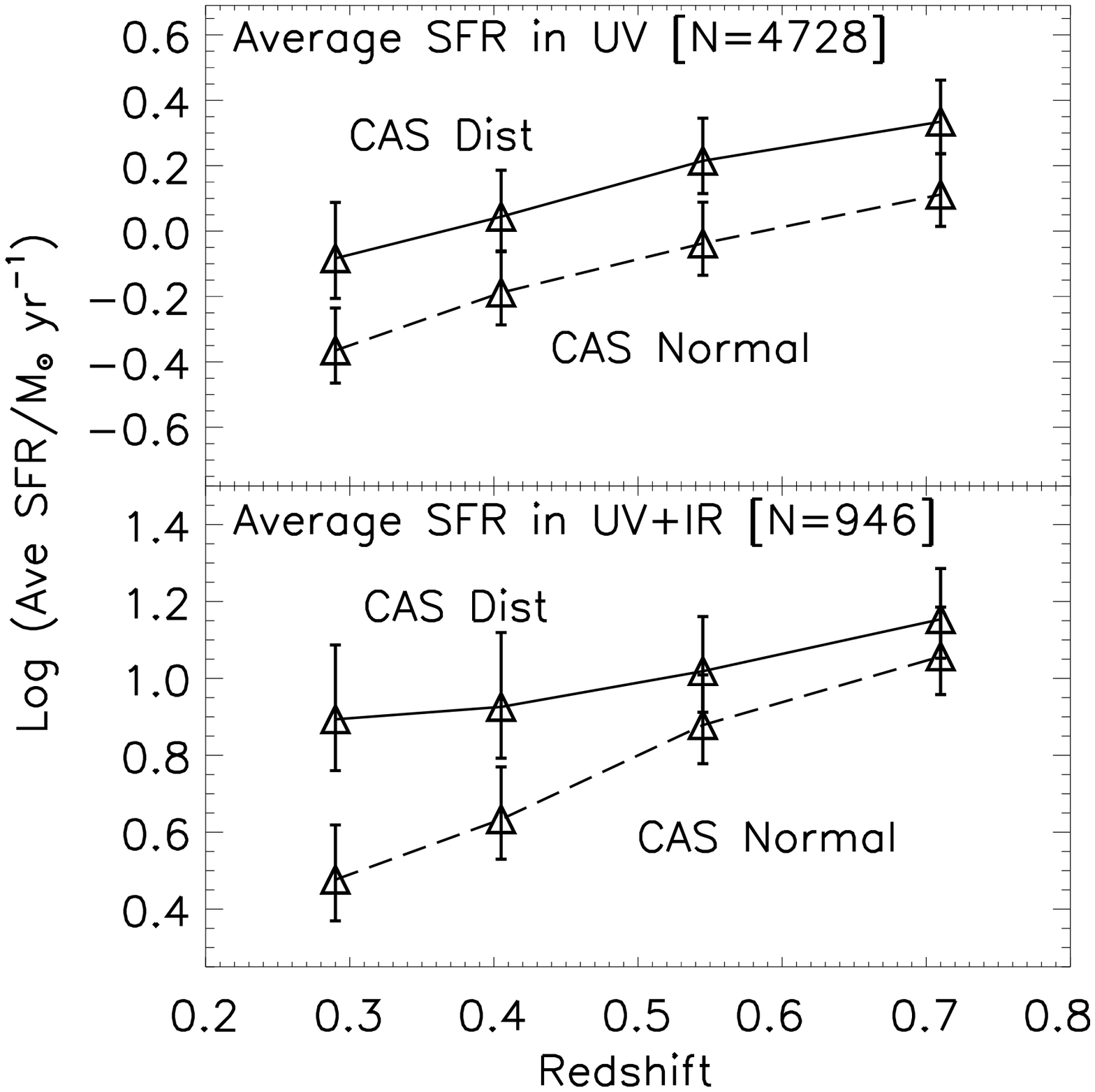}{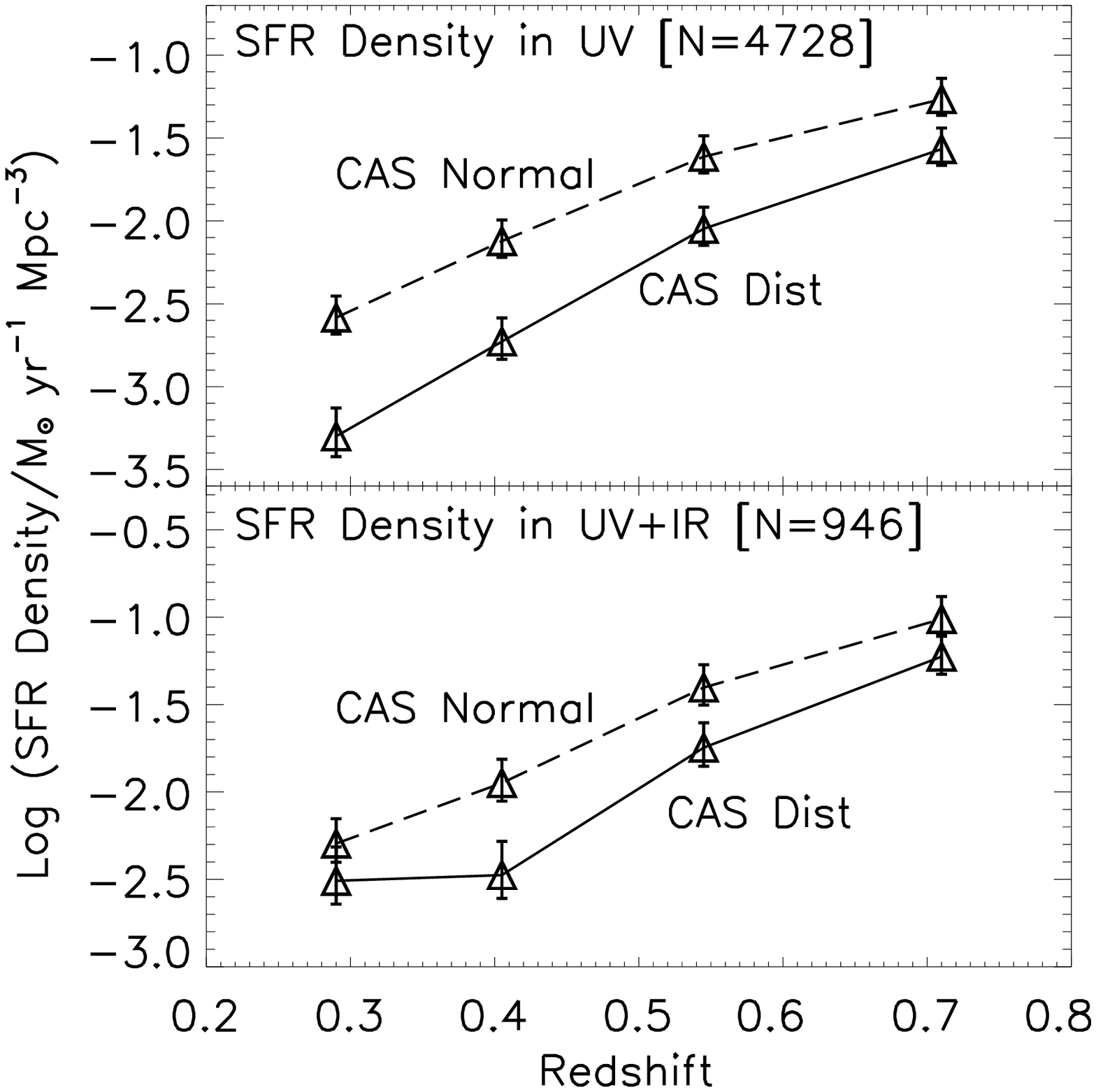}
\caption{The average star formation rate ({\itshape Left}) and the cosmic star formation rate 
density ({\itshape Right}) as functions of redshift between the systems that the CAS criteria 
[$\,A>S$ and $\,A>0.35$] identify as normal and identify as distorted/interacting systems. 
According to the CAS criteria, the severe decline in the cosmic SFR density we observe between 
$z\sim$~0.2--0.8 (Madau et al. 1996) is driven primarily by the shutdown of star formation in 
normal undisturbed systems rather than distorted/interacting systems.}
\end{figure}

\begin{figure}[!ht]
\plotfiddle{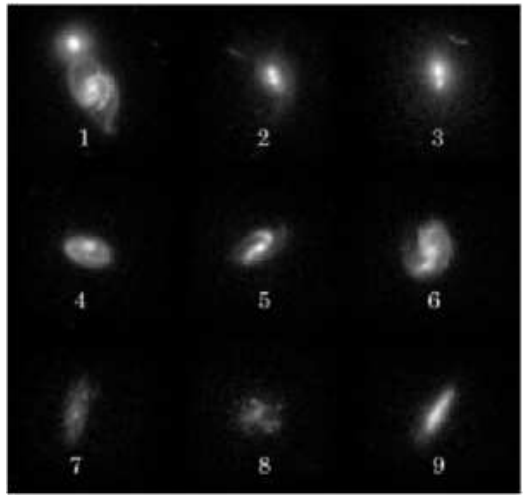}{160 pt}{0}{90}{90}{-110 pt}{-20 pt}
\caption{Galaxies 1-3: Visually classified as strongly distorted/interacting, but CAS criteria identify as normal.
         Galaxies 4-9: Visually classified as normal types, but CAS criteria identify as distorted/interacting.}
\vspace{-0.5 cm}
\end{figure}
\vspace{-0.5 cm}
\section{Conclusions}

\begin{enumerate}

\item 
The CAS criteria ($\,A>S$ and $\,A>0.35$) captures 38\% to 58\% (Fig.~1) of  
the galaxies visually classified as interacting.  This is not unexpected,
given that on average, the $A >$~0.35 criterion is only satisfied for 
one third of the major merger timescale in N-body simulations 
(Conselice 2006).

\item We inspected the morphologically distorted galaxies, which the CAS 
merger criteria fail to capture and find the following. 
The criteria do not pick up systems where the
externally-triggered tidal features (e.g., light bridges between galaxies, 
tidal tails, arcs, shells, ripples) and tidal debris
(e.g., small accreted satellite in the main disk  of a galaxy), 
contribute  less than $35\%$ of the total light 
(e.g., Fig.~3, cases 1 and 3). In addition, in galaxies with close 
double nuclei where the center  is assumed to be between the two nuclei, the 
resulting low $A$ value will prevent the system from satisfying 
the CAS criteria (e.g., Fig.~3, case 2).

\item The CAS criteria suffer from contamination by relatively undistorted, so-called `normal' 
galaxies. The contamination rate (i.e. fraction of `normal' galaxies misidentified by CAS as 
distorted/interacting) increases with redshift (Fig.~1). This is due to several factors:
\begin{enumerate}
\item S (clumpiness) and A (asymmetry) values are higher in systems with increased star formation. 
Small-scale asymmetries due to stochastic star formation can be separated from truly interacting
 systems by visual classification. However, CAS (designed in rest-frame optical) 
may identify high star-forming systems as interacting 
(e.g., Fig.~3, cases 4 and 6) at bluer/UV wavelengths as in $z\sim$~0.6--0.8. (Fig.~1)
\item Galaxies whose outer parts look irregular (e.g., Fig.~3, cases 7 and 8),  
whether intrinsically or due to cosmological surface brightness dimming, 
can be picked by the CAS criteria.
\item In systems without a clear center (e.g. dusty bright galaxies, local and distant Irrs, 
and distant Sds that are dimmed to resemble Irrs) the A (asymmetry) parameter can be high (e.g., 
Fig.~3 cases 4 and 8).
\item In edge-on systems and compact systems, where the light profile is steep, small centering 
inaccuracies can lead to large A (asymmetry) values. (e.g., Fig.~3 case 9) 
\end{enumerate}

\item In Jogee, Miller, Penner, et al. (2008), we reported 
the following from visual classification. The average  SFR of 
strongly disturbed/interacting systems is only modestly enhanced, 
by a factor of 2--3,  with respect to  normal undisturbed galaxies. 
Contrary to common lore, large order of magnitude enhancements
in the SFR are rare in strongly disturbed systems. In fact, such 
systems contribute  below $20\%$ of the  cosmic SFR density over 
$z\sim$~0.2--0.8. Thus, the decline over this regime is driven 
primarily by a shutdown in the star formation of 
relatively undisturbed galaxies. Our results are consistent 
with earlier findings, over a narrower redshfit bin ($z\sim$~0.65--0.75), 
that relatively undisturbed galaxies produce  most of the UV   
(Wolf et al. 2005) and IR (Bell et al. 2005)  luminosity density.
\end{enumerate}

\vspace{-0.1 cm}

A certain element of uncertainty and subjectivity is inherent in  
visual classification. However, the robustness of the results can 
be illustrated by an independent analysis based on the 
CAS system, yielding the same conclusions. In particular, we show 
that the cosmic SFR density is dominated by  systems classified 
as `Normal' by CAS,  rather than as  distorted  (Fig.~2).
The interpretation here is complicated due to the fact that 
the CAS `Normal' class includes a small number of distorted galaxies 
missed by the CAS merger criteria, while  the CAS `Distorted' class
is contaminated by a number of Irregular systems. Nonetheless, it 
is encouraging that the same conclusions are reached with CAS.

\vspace{-0.0 cm}
\acknowledgements
We acknowledge support from NSF grand AST-0607748, NASA LTSA grant 
NAG5-13063, as well as HST grants GO-9500 from STScI, which is 
operated by AURA, Inc., for NASA, under NAS5-26555.


\vspace{-0.25 cm}
\begin{thebibliography}{}
\bibitem[Bell et al.(2005)]{2005ApJ...625...23B} Bell, E.~F., et al.\ 2005, 
\apj, 625, 23 
\bibitem[Bershady et al.(2000)]{Bers2000} Bershady, M.~A., 
Jangren, A., \& Conselice, C.~J.\ 2000, \aj, 119, 2645 
\bibitem[Conselice et al.(2000)]{Cons2000} 
Conselice, C.~J.,  Bershady, M.~A., \& Jangren, A.\ 2000, \apj, 529, 886 
\bibitem[Conselice et al.(2003)]{Cons2003} 
Conselice, C.~J., Bershady, M.~A., Dickinson, M., \& Papovich, C.\ 2003, \aj, 126, 1183
\bibitem[]{} 
Conselice, C.~J.\ 2006,\apj, 638, 686
\bibitem[Jogee, S. et al. (2008)]{Joge2008}Jogee et al. 2008, Proceedings of 
"Formation and Evolution of Galaxy Disks", held in Rome, 1-5 October 2007  (arXiv:0802.3901)
\bibitem[Madau et al.(1996)]{Mada1996} 
Madau, P., et al. 1996, \mnras, 283, 1388 
\bibitem[Wolf et al.(2005)]{2005ApJ...630..771W} Wolf, C., et al.\ 2005, 
\apj, 630, 771 

\end{thebibliography}
\end{document}